\documentclass[a4paper,fleqn,usenatbib,useAMS]{mnras}

\usepackage{graphicx}	% Including figure files
\usepackage{amsmath}	% Advanced maths commands
\usepackage{amssymb}	% Extra maths symbols
\usepackage{multicol}        % Multi-column entries in tables
\usepackage{multirow}
\usepackage{bm}		% Bold maths symbols, including upright Greek
\usepackage{pdflscape}	% Landscape pages
\usepackage{slashed}
\usepackage{lipsum}
\usepackage{nicematrix}
\usepackage{bbold}
\usepackage{dsfont}
\usepackage{pbox}
\usepackage{scalerel,stackengine}
\stackMath
\newcommand\reallywidehat[1]{%
\savestack{\tmpbox}{\stretchto{%
  \scaleto{%
    \scalerel*[\widthof{\ensuremath{#1}}]{\kern-.6pt\bigwedge\kern-.6pt}%
    {\rule[-\textheight/2]{1ex}{\textheight}}%WIDTH-LIMITED BIG WEDGE
  }{\textheight}% 
}{0.5ex}}%
\stackon[1pt]{#1}{\tmpbox}%
}
\DeclareTextFontCommand{\emph}{\em}
\newcommand{\mmat}[1]{\mathsf{#1}}
\newcommand{\mymat}[1]{\mmat{#1}}
\usepackage{dsfont}
\usepackage{cancel}
\usepackage{xcolor}

\newcommand{\mua}[1]{\boldsymbol{\mu}_{,#1}}
\newcommand{\muab}[2]{\boldsymbol{\mu}_{,#1#2}}
\newcommand{\maMmb}[2]{\mua{#1}^{\mathrm{t}}\mymat{C}^{-1}\mua{#2}}

\newcommand{\maMmbc}[3]{\mua{#1}^{\mathrm{t}}\mymat{C}^{-1}\muab{#2}{#3}}
\newcommand{\mabMmcd}[4]{\muab{#1}{#2}^{\mathrm{t}}\mymat{C}^{-1}\muab{#3}{#4}}
\newcommand{\Dp}[1]{\Delta_{#1}}

\newcommand{\mat}[3]{\mymat{#1}_{\scriptscriptstyle {#2}\times{#3}}}
\usepackage[T1]{fontenc}
\usepackage{ae,aecompl}
\usepackage{arydshln}

\usepackage{newtxtext,newtxmath}
\defcitealias{DALI}{S14}
\defcitealias{DALII}{S15}
\title[Expensive likelihoods in precision cosmology]{Extremely expensive likelihoods: A variational-Bayes solution for precision cosmology}

\author[Rizzato \& Sellentin]{
Matteo Rizzato,$^{1}$\thanks{Contact e-mail: sellentin@strw.leidenuniv.nl}
Elena Sellentin$^{1,2}$
\\
$^{1}$Leiden Observatory, Leiden University, Huygens Laboratory, Niels Bohrweg 2, NL-2333 CA Leiden, The Netherlands.\\
$^{2}$Mathematical Institute, Leiden University, Snellius Building, Niels Bohrweg 1, NL-2333 CA Leiden, The Netherlands.
}

\date{Last updated ; in original form }

\pubyear{2022}

\begin{document}
\label{firstpage}
\pagerange{\pageref{firstpage}--\pageref{lastpage}}
\maketitle

% Abstract of the paper
\begin{abstract}
We present a variational-Bayes solution to compute non-Gaussian posteriors from extremely expensive likelihoods. Our approach is an alternative for parameter inference when MCMC sampling is numerically prohibitive or conceptually unfeasible. For example, when either the likelihood or the theoretical model cannot be evaluated at arbitrary parameter values, but only previously selected values, then traditional MCMC sampling is impossible, whereas our variational-Bayes solution still succeeds in estimating the full posterior. In cosmology, this occurs e.g.~when the parametric model is based on costly simulations that were run for previously selected input parameters.  We demonstrate the applicability of our posterior construction on the KiDS-450 weak lensing analysis, where we reconstruct the original KiDS MCMC posterior at 0.6\% of its former numerical posterior evaluations. The reduction in numerical cost implies that systematic effects which formerly exhausted the numerical budget could now be included. 
\end{abstract}

% Select between one and six entries from the list of approved keywords.
% Don't make up new ones.

%%%%%%%%%%%%%%%%%%%%%%%%%%%%%%%%%%%%%%%%%%%%%%%%%%

%%%%%%%%%%%%%%%%% BODY OF PAPER %%%%%%%%%%%%%%%%%%

\section{Introduction}

\label{Intro}
Modern cosmological surveys are designed to probe the evolved Universe with superb data precision. These surveys scout for faint signatures of new physics on top of the standard model of cosmology, or even a breakdown of the standard model itself. The precision of upcoming cosmological data sets promises the most competitive constraints on the cosmic law of gravity, the equation of state of dark energy and dark matter \citep{RuthRoy}. Within the framework of a cosmological model, these data sets are also used to constrain the number of neutrino species and the neutrino masses \citep{Steffen, RuthNeutrinos, JulienForecastNeutrinos}. 

During the next decade, the forefront of cosmological surveys will be composed of the Euclid satellite 
\citep{2011arXiv1110.3193L} and the ground-based `Legacy Survey of Space and Time' \citep{2009arXiv0912.0201L}. These surveys take over from a successful series of contemporary surveys, such as Kilo Degree Survey \citep{HSC}, the Dark Energy Survey \citep{DESnoses} and the Hyper Suprime Cam on Subaru \citep{HSC}. A major challenge for these surveys is that the improvement in data quality has to be matched by an equal improvement in data analysis accuracy: insufficient data analysis techniques will lose information actually contained in the data, and introduce incorrect assumptions, thereby causing biases and uncertainties larger than nominally quoted, and certainly larger than those resulting from more adequate analysis techniques.

In consequence, the field of cosmology is currently entering a phase where theoretical modelling of the data and likelihood calculations both become prohibitively expensive to compute. For this paper, we therefore assume that the aim is to infer parameters from data which are so precise or so complex that only few model evaluations at the required precision are possible. In particular, we assume that it is impossible to evaluate the theoretical model and the likelihood $\approx 10^5-10^6$ times, which is a typical number of evaluations employed by current cosmological analyses using Monte Carlo Markov Chains (MCMC). 

In the most extreme case, parameter inference will have to start from e.g.~50 highly accurate simulations run for different cosmological models. In this case, not only is the number of possible posterior evaluations fixed to 50, but the positions where the posterior can be evaluated are additionally pre-determined: it can only be computed at the parameter values chosen as input parameters for the simulations. In contrast, an MCMC run would require that the posterior can be evaluated at any arbitrary position throughout parameter space. 

Put briefly, we are searching for a posterior construction technique that succeeds despite computing the posterior at few, potentially pre-determined fixed parameter values.

To this end, we propose in this paper a variational-Bayes solution, which fits a flexible distribution to few evaluations of the likelihood, thereby yielding an approximate posterior. The essential assumption enabling the speed-up is that the posterior be fairly smooth, though not necessarily unimodal. 

Our paper is structured as follows. In Sec.~\ref{VarBays}, we present the class of variational-Bayes methods. These methods fit a parametric distribution to evaluations of the posterior at few points. Accordingly, the quality of the resulting fit is determined by a suitable choice of the flexible distribution. In Sec. \ref{MainSec}, we therefore set up our own distribution to be fitted. The result from our variational-Bayes technique is compared in Sec. \ref{TestingMethod} to the publicly available MCMC from the KiDS-450 tomographic weak gravitational lensing analysis \citep{2017MNRAS.465.1454H}. We finally summarize the benefits of the method while paving the way to follow up studies in Sec.~\ref{discANDconcl}. In spirit, our variational-Bayes approach is similar to \citet{Luca}, which uses a different (sometimes divergent) variational distribution.

\section{Variational-Bayes methods}
\label{VarBays}
Variational-Bayes methods construct a posterior by fitting a candidate distribution with free shape parameters to a few calls of the target posterior \citep{10.5555/971143}.

In this paper, we denote the model parameters of physical interest by $\bm{p}$ and the data by $\bm{x}$, the sought posterior distribution of the parameters as constrained by the data is  $\mathcal{P}(\bm{p}|\bm{x})$. Bayes theorem relates $\mathcal{P}(\bm{p}|\bm{x})$ to the likelihood $\mathcal{L}(\bm{x}|\bm{p})$ and to the parameters' prior $\pi(\bm{p})$ as \begin{equation}
\label{Bayes}
    \mathcal{P}(\bm{p}|\bm{x})\propto \mathcal{L}(\bm{x}|\bm{p})\pi(\bm{p}).
\end{equation}
The aim of parameter inference is to calculate this posterior, usually quoting the maximum a posteriori (MAP) estimate of best-fitting parameters and plotting the marginal distributions of parameter pairs.  For precision cosmology with expensive likelihoods, the usual MCMC sampling to this end can become unfeasible. 

Variational-Bayes methods can compute posteriors at a lower numerical cost as they provide an analytical approximation to the posterior. The approximation is computed based on a few posterior evaluations only. The approximating distribution is usually known as the \textit{variational distribution} $\mathcal{V}\left(\bm{p}|\bm{x},\bm{P}\right)$ and it is parametrized by a set of shape parameters $\bm{P}$, also referred to as `variational parameters', whose values have to be determined such that the variational distribution approximates the wanted posterior in the sense of
\begin{equation}
 \mathcal{P}\left(\bm{p}|\bm{x}\right) \approx   \mathcal{V}(\bm{p}|\bm{x},\hat{\bm{P}}),  
 \label{approx}
\end{equation}
where $\hat{\bm{P}}$ are the best-fitting shape parameters. For precision cosmology, it is obviously crucial that the posterior approximation indicated in Eq.~(\ref{approx}) is of convincing quality. Achieving this quality constitutes the primary challenge to be addressed in this paper.

Technically, the key to a successful variational-Bayes solution is a variational distribution that is so flexible in shape that it can accurately represent arbitrarily formed posteriors. Accordingly, we develop in this paper a variational-Bayes method where the  variational distribution $\mathcal{V}\left(\bm{p}|\bm{x},\bm{P}\right)$ is a generalization of the non-Gaussian `DALI' distribution presented in \cite{DALI} (hereafter S14) and in  \cite{DALII} (hereafter S15). DALI distributions form a series and are positive definite and non-divergent at any order. They are based on a Taylor expansion, and can hence be proven to converge towards the true distribution. This Taylor expansion can be truncated at any order, but in practice truncating at the second order is almost always sufficient \citep{WBjoern}. Accordingly, we will here build upon the second-order DALI approximation obtained in Eq.~(15) of \citetalias{DALI}. The adoption of DALI is a deliberate choice: Fitting any arbitrary multivariate polynomial instead can lead to regions in parameter space where the posterior probability takes (meaningless) negative values, or regions of infinite probability, far away from the peak. DALI avoids these problems by construction as detailed in \citetalias{DALI}.

\section{Setting up the variational distribution}
\label{MainSec}
In this section, we set up the variational distribution $\mathcal{V}\left(\bm{p}|\bm{x},\bm{P}\right)$ whose shape parameters $\bm{P}$ will be fitted to a few posterior samples. We start by reviewing in Sec.~\ref{Dali1} the DALI formalism as developed in \citetalias{DALI} and \citetalias{DALII}, and then discuss how to optimize it for use as a variational distribution in Sec.~\ref{FittingFormalism}. This derivation is somewhat involved and readers might want to skip directly to Sec.~\ref{TestingMethod} where the performance of the posterior reconstruction technique is demonstrated on the KiDS-450 data.

\subsection{DALI review}
\label{Dali1}
We shortly review the DALI formalism as published in \citetalias{DALI} and \citetalias{DALII}. 

Let us consider a Gaussian data set $\bm{x}$ with expectation value $\bm{\mu}$ and covariance matrix $\mymat{C}$.
The parameters to be inferred are $\bm{p}$, and the mean $\bm{\mu}(\bm{p})$ is to be determined. For this paper it suffices to use a constant $\mymat{C}$, for parameter dependent $\mymat{C}$ see \citetalias{DALII}.

DALI forecasts the posterior of parameters $\bm{p}$, i.e.~it predicts the posterior \emph{before} actual data are taken. To achieve these forecasts, DALI follows the line of thought of Fisher matrix forecasts \citep{Tegmark:1996bz} by computing the expected log-posterior by averaging out the not yet collected data. Accordingly, both DALI and Fisher matrix forecasts compute
\begin{equation}
    \mathcal{P}(\bm{p}) \propto \mathrm{exp}\langle\mathrm{log}\ \mathcal{P}(\bm{p}|\bm{x})\rangle,
    \label{avpost}
\end{equation}
where the angular brackets denote the average over the data $\bm{x}$. While Fisher forecasts are limited to Gaussian posterior approximations, DALI yields the actual non-Gaussian posteriors. 

We now define the deviation from the maximum a posteriori point (MAP) $\hat{\bm{p}}$ as
\begin{equation}
\bm{\Delta}= \bm{p} - \hat{\bm{p}}.
\end{equation}
The DALI formalism at second order provides the following series expansion $\mathcal{D}\left(\bm{p}\right)$ to the average parameter posterior in Eq.~(\ref{avpost}):
\begin{align}
    \mathcal{D}\left(\bm{p}\right) \propto \text{exp}\Big[ &-\frac{1}{2}\maMmb{\alpha}{\beta}\ \Dp{\alpha}\Dp{\beta} \nonumber \\
    &- \frac{1}{2}\maMmbc{\alpha}{\beta}{\gamma}\Dp{\alpha}\Dp{\beta}\ \Dp{\gamma}\nonumber\\
    &- \frac{1}{8}\mabMmcd{\alpha}{\beta}{\gamma}{\delta}\ \Dp{\alpha}\Dp{\beta}\Dp{\gamma}\Dp{\delta} \Big].
    \label{olddali}
\end{align}
Partial derivatives are indicated by commas, ${\bm{\mu}_{,\alpha}\equiv\partial\bm{\mu}/\partial p_{\alpha}}$, ${\bm{\mu}_{,\alpha\beta}\equiv\partial^2\bm{\mu}/\partial p_{\alpha}\partial p_{\beta}}$ and we imply Einsteinian summation over repeated indices. Note that the first summand in Eq.~(\ref{olddali}) corresponds to the Fisher approximation. The remaining two summands appear simultaneously in the second order DALI approximation and jointly guarantee positive definiteness, finiteness, and convergence towards the true posterior.

In this paper, Greek indices run from $1$ to $d=\text{dim}\left[\bm{p}\right]$. The dimension of the data is $D=\text{dim}\left[\bm{x}\right]$. 
We also introduce calligraphic indices, such as $\mathcal{I}$, in order to denote compound indices for ordered couples $(\alpha\beta)$ within the list
\begin{align}
      \big\{(11),(12),\dots,(\alpha\beta)_{\beta\geq \alpha},\dots,(dd)\big\}.
\end{align}
Therefore, calligraphic indices take values ${\mathcal{I} = 1,\dots,s}$ and $s \equiv d (d+1)/2$.

\subsection{Variational Inference with DALI}
\label{FittingFormalism}
DALI, as reviewed in Sec.~\ref{Dali1}, handles parameter forecasts where data have not yet been collected and hence, the parameters are not yet \textit{inferred}. Variational inference does, however, handle actual data and actual inference. In this section, we therefore modify the DALI formalism for use in variational inference, such that actual data can be used. In other words, the aim is to modify $\mathcal{D}(\bm{p})$ of  Eq.~(\ref{olddali}) such that it can be used as variational distribution
 $\mathcal{V}\left(\bm{p}|\bm{x},\bm{P}\right)$.

We begin by noting that the components of $\bm{\mu}_{\alpha}, \bm{\mu}_{\alpha\beta}$ and $\mymat{C}$ in Eq.~(\ref{olddali}) carry physical units and have hence (conventional but otherwise) arbitrary values. We remove this arbitrariness by whitening the data, the advantage being that this whitening eliminates $\mymat{C}$ as degenerate free shape parameters we would otherwise need to fit for.

 From this point on, we will often implicitly imply the dependence on the measured data set $\bm{x}$ in order to simplify the notation.

We introduce the Cholesky decomposition of $\mymat{C}^{-1}$ in terms of a lower triangular matrix $\mymat{L}$ as $\mymat{L}\mymat{L}^{\mathrm{t}} = \mymat{C}^{-1}$ and define the vectors
\begin{equation}
\label{MandV}
    \bm{V}_{\alpha} \equiv \mymat{L}^{\mathrm{t}}\boldsymbol{\mu}_{,\alpha},\quad  \bm{M}_{\alpha\beta} \equiv \mymat{L}^{\mathrm{t}}\boldsymbol{\mu}_{,\alpha\beta},
\end{equation}
and write Eq.~(\ref{olddali})  as 
\begin{align}
    \nonumber\mathcal{V}\left(\bm{p}|\bm{P}\right) \propto \ \ \text{exp}\Big[
    &-\frac{1}{2} \bm{V}_{\alpha}^{\mathrm{t}}\bm{V}_{\beta}\ \Dp{\alpha}\Dp{\beta}\\ 
    &- \frac{1}{2}\ \bm{V}_{\alpha}^{\mathrm{t}}\bm{M}_{\beta\gamma}\ \Dp{\alpha}\Dp{\beta}\Dp{\gamma}\nonumber\\
    &- \frac{1}{8} \bm{M}_{\alpha\beta}^{\mathrm{t}}\bm{M}_{\gamma\delta}\ \Dp{\alpha}\Dp{\beta}\Dp{\gamma}\Dp{\delta}\Big].\label{re_written1}
\end{align}
The scalar products between the vectors $\bm{V}_{\alpha}$ and  $\bm{M}_{\alpha\beta}$ are invariant under the orthogonal group: while any orthogonal matrix simultaneously applied to these vectors would affect their components, their scalar products will remain invariant. Hence, we remove the dependence on the vector orientation by  expressing the scalar products in terms of the vector lengths  $V_{\alpha}$, $M_{\alpha\beta}$ and of the cosine of the angle between them, as follows:
\begin{align}
&\bm{V}_{\alpha}^{\mathrm{t}}\bm{V}_{\beta} = V_{\alpha}V_{\beta}\cos\left(\theta_{\alpha\beta}\right), \label{OldIndex0}\\
&\bm{V}_{\alpha}^{\mathrm{t}}\bm{M}_{\beta\gamma} = V_{\alpha}M_{\beta\gamma} \cos\left(\theta_{\alpha;\beta\gamma}\right),\label{OldIndex1}\\
&\bm{M}_{\alpha\beta}^{\mathrm{t}}\bm{M}_{\gamma\delta} = M_{\alpha\beta}M_{\gamma\delta}\cos\left(\theta_{\alpha\beta;\gamma\delta}\right).\label{OldIndex2}
\end{align}

The aim is then to construct a variational distribution  $\mathcal{V}\left(\bm{p}|\bm{x},\bm{P}\right)$ such that
\begin{align}
\label{exponent1}
    &\mathcal{V}\left(\bm{p}|\bm{x},\bm{P}\right)\propto \mathrm{exp}\left[-\mathcal{Y}(\bm{\Delta},\bm{x},\bm{P})\right],\\ \label{exponent2}
&\mathcal{Y}\left(\bm{\Delta},\bm{x},\bm{P}\right) = \frac{1}{2}\bm{B}^{\mathrm{t}}\left(\bm{\Delta},\bm{x},\bm{P}\right)\bm{B}\left(\bm{\Delta},\bm{x},\bm{P}\right),
\end{align}
where $\mathcal{Y}(\bm{\Delta},\bm{x},\bm{P})$ has to be a quadratic form in order to guarantee a normalizable, positive definite posterior.

To advance towards the definition of a vector $\bm{B}$ as in Eq.~(\ref{exponent2}), we define the vectors $\boldsymbol{\Delta}^{\mathrm{v}}$ and $\boldsymbol{\Delta}^{\mathrm{m}}$ as
\begin{align}
    &\boldsymbol{\Delta}^{\mathrm{v}} \equiv \{\Dp{1}V_{1},\dots,\Dp{d}V_{d} \},\\
    &\boldsymbol{\Delta}^{\mathrm{m}} \equiv \{\Dp{1}\Dp{1}M_{1}, 2\Dp{1}\Dp{2}M_{2}, \dots,\nonumber\\  &\hspace{2cm}2\left(1-\delta_{\alpha\beta}\right)\Delta_{\alpha}\Delta_{\beta}M_{(\alpha\beta)}, \dots, \Dp{d}\Dp{d}M_{s}\},
    \label{vecsis}
\end{align}
and the matrices $\mymat{A}^{\mathrm{v}},\ \mymat{A}^{\mathrm{vm}}$ and $\mymat{A}^{\mathrm{m}}$, whose components are 
\begin{align}
\label{Matrices1}
      A^{\mathrm{v}}_{\alpha\beta} &\equiv \cos\left(\theta_{\alpha\beta}\right),\\\label{Matrices2}
      A^{\mathrm{v}\mathrm{m}}_{\alpha \mathcal{I}} &\equiv \cos\left(\theta_{\alpha;\mathcal{I}}\right),\\\label{Matrices3}
      A^{\mathrm{m}}_{\mathcal{I} \mathcal{J}} &\equiv \cos\left(\theta_{\mathcal{I};\mathcal{J}}\right).
\end{align}
In this  notation, the negative logarithm of the variational distribution from Eq.~(\ref{exponent1}) is
\begin{equation}
\label{Loss3}
    \mathcal{Y}(\bm{\Delta},\bm{P}) = 
    \frac{1}{2}{\left(\boldsymbol{\Delta}^{\mathrm{v}}\right)}^{\mathrm{t}}\mathrm{\textbf{A}}^{\mathrm{v}}\boldsymbol{\Delta}^{\mathrm{v}} +\frac{1}{2}\   {\left(\boldsymbol{\Delta}^{\mathrm{v}}\right)}^{\mathrm{t}} \mathrm{\textbf{A}}^{\mathrm{v}\mathrm{m}}\boldsymbol{\Delta}^{\mathrm{m}} + \frac{1}{8} {\left(\boldsymbol{\Delta}^{\mathrm{m}}\right)}^{\mathrm{t}} \mathrm{\textbf{A}}^{\mathrm{m}}\boldsymbol{\Delta}^{\mathrm{m}}. 
\end{equation}

However, Eq.~(\ref{Loss3}) is not automatically a quadratic form, if the matrices $\mymat{A}^{\mathrm{v}},\ \mymat{A}^{\mathrm{vm}}$ and $\mymat{A}^{\mathrm{m}}$ are independent of each other. Accordingly, the necessary relation between these matrices still has to be enforced, such that Eq.~(\ref{Loss3}) is a valid candidate for our log-posterior.  
We hence relate these matrices to each other by jointly decomposing them as follows.

The Fisher matrix $\mymat{F}$ is
\begin{equation}
\label{FromFtoAv}
   F_{\alpha\beta} =  \bm{V}_{\alpha}^{\mathrm{t}}\bm{V}_{\beta} = V_{\alpha}V_{\beta}\cos\left(\theta_{\alpha\beta}\right) = V_{\alpha}V_{\beta}A^{\mathrm{v}}_{\alpha\beta}, 
\end{equation}
and has to be positive definite. Consequently,  $\mymat{A}^{\mathrm{v}}$ has to be positive definite as well. We therefore Cholesky-decompose it
\begin{equation}
\label{CholeskyAv}
    \mymat{A}^{\mathrm{v}} = \mymat{L}^{\mathrm{v}}\left(\mymat{L}^{\mathrm{v}}\right)^{\mathrm{t}},\quad
    L^{\mathrm{v}}_{\alpha\beta>\alpha} = 0,\  L^{\mathrm{v}}_{\alpha\alpha} \gneq 0.
\end{equation}
Given the scalar products of Eq.~(\ref{Matrices1}), we have the constraint  
\begin{equation}
\label{Prior1}
%    -1 \lneq A^{\mathrm{v}}_{\alpha\beta,\beta\neq\alpha} \lneq +1,
\quad A^{\mathrm{v}}_{\alpha\alpha} = 1.
\end{equation}
To enforce this condition, we request the rows $\{\bm{v}_{\alpha}\}$ of the matrix $\mymat{L}^{\mathrm{v}}$ to be vectors of unitary norm. 

As for $\mymat{A}^{\mathrm{m}}$, we point out that if it is positive definite, then we will have
\begin{equation}
\begin{aligned}
    \mymat{A}^{\mathrm{m}}\ \text{positive definite} &\leftrightarrow \forall \bm{\Delta}^{\mathrm{m}} \in \mathbb{R}^{s},\ \left({\bm{\Delta}^{\mathrm{m}} }\right)^{\mathrm{t}} \mymat{A}^{\mathrm{m}} \bm{\Delta}^{\mathrm{m}} \gneq 0.\\
\end{aligned}
\end{equation}
Hence, we see that $\mymat{A}^{\mathrm{m}}$ must indeed be positive definite, otherwise the scalar product with $\bm{\Delta}^{\mathrm{m}}$ will pick up a minus sign, which will cause the posterior to diverge at large distances from the peak. We therefore Cholesky-decompose it too, yielding
\begin{equation}
\label{CholeskyM}
    \mymat{A}^{\mathrm{m}} \equiv \mymat{L}^{\mathrm{m}}\left({\mymat{L}^{\mathrm{m}}}\right)^{\mathrm{t}}.\quad
\end{equation}
This allows us to implement the prior constraints
\begin{equation}
        L^{\mathrm{m}}_{\mathcal{I}\mathcal{J}>\mathcal{I}} = 0,\quad L^{\mathrm{m}}_{\mathcal{I}\mathcal{I}} > 0.
\end{equation}
Additionally, we again demand the rows $\{\bm{m}_{\mathcal{I}}\}$ of $\mymat{L}^{\mathrm{m}}$ to be vectors of unitary norm, which ensures that 
\begin{equation}
\label{Prior2}
A^{\mathrm{m}}_{\mathcal{I}\mathcal{I}} = 1.
\end{equation}

We now turn to the decomposition of $\mymat{A}^{\mathrm{vm}}$ which is a rectangular matrix. Decomposing this matrix correctly is crucial to yield a positive-definite and normalizable posterior. We hence review the QR-decomposition of rectangular matrices in App.~(\ref{App:qr}).

We introduce rectangular orthogonal matrices $\mymat{Q}\in\mathbf{R}^{n\times p}$, $n\gneq p$, which satisfy the relation $\mymat{Q}^{\mathrm{t}}\mymat{Q} = \mat{1}{p}{p}$. For $n=p$, we obtain the elements of the orthogonal group $\mymat{O}\left(n\right)$. Furthermore, we define the following matrices:
\begin{equation}
\label{Usefulnotation}
    \mat{1}{d_1}{d_2} \in \mathbb{R}^{d_1\times d_2},\quad d_1 \geq d_2,\quad \left[\mat{1}{d_1}{d_2}\right]_{ij}\equiv \begin{cases}
                             \delta_{ij}&\text{for}\ i\leq d_1,\\
                             0          &\text{elsewhere}.\\
                            \end{cases}
\end{equation}
Matrices defined by Eq.~(\ref{Usefulnotation}) will be the usual identity matrices only if square. If rectangular, additional zero-rows appear, e.g. we have 
\begin{equation}
    \mat{1}{3}{2} = \begin{pmatrix}
                    1 & 0\\
                    0 & 1\\
                    0 & 0\\
                    \end{pmatrix}.
\end{equation}

We now introduce an auxiliary dimension $D$ and set $D\geq s\gneq d$: this allows the $s$ vectors $\{\bm{M}_{\mathcal{I}}\}$ to be linearly independent, such that the matrix of their relative angles $\mymat{A}^{\mathrm{m}}$ is as general as can be.
We gather the $d$ vectors $\{\bm{V}_{\alpha}\}$ as the columns of a matrix ${\mymat{V}\in\mathbb{R}^{\scriptstyle D\times d}}$ that we  QR-decompose as
\begin{equation}
\label{vqr}
    \mymat{V} = \mymat{Q}^{\mathrm{v}}\ \mymat{R}^{\mathrm{v}},\quad \mymat{Q}^{\mathrm{v}}\in\mathrm{O}\left( D\times d\right),\ 
                                                                   \mymat{R}^{\mathrm{v}}\in\mathbb{R}^{\scriptstyle d\times d},
\end{equation}
$\mymat{R}^{\mathrm{v}}$ being upper triangular.
This decomposition is unique whenever $\mymat{R}^{\mathrm{v}}_{ii}>0$.
If we define the normalization matrices $\mymat{N}^{\mathrm{v}}$, $\mymat{N}^{\mathrm{m}}$ as ${\mymat{N}^{\mathrm{v}}_{ij}=\delta_{ji}/V_i}$ and $\mymat{N}^{\mathrm{m}}_{\mathcal{I}\mathcal{J}}=\delta_{\mathcal{I}\mathcal{J}}/M_{\mathcal{I}}$, respectively, we can then write the matrix $\mymat{A}^{\mathrm{v}}$ from Eq.~(\ref{Matrices1}) as \begin{align} 
\label{VectorisedAV}
    &\mymat{A}^{\mathrm{v}} = \mymat{N}^{\mathrm{v}}\ \mymat{V}^{\mathrm{t}}\ \mymat{V}\ \mymat{N}^{\mathrm{v}} = \mymat{N}^{\mathrm{v}}\ \left(\mymat{R}^{\mathrm{v}}\right)^{\mathrm{t}}\left(\mymat{Q}^{\mathrm{v}}\right)^{\mathrm{t}}\ \mymat{Q}^{\mathrm{v}}\mymat{R}^{\mathrm{v}}\ \mymat{N}^{\mathrm{v}} = \left(\cancel{\mymat{R}}^{\mathrm{v}}\right)^{\mathrm{t}}\cancel{\mymat{R}}^{\mathrm{v}},\\
    &\cancel{\mymat{R}}^{\mathrm{v}}\equiv\mymat{R}^{\mathrm{v}}\mymat{N}^{\mathrm{v}}.
\end{align}
However, since the Cholesky decomposition for a positive definite matrix is unique, we conclude, by comparing Eq.~(\ref{CholeskyAv}) and Eq.~(\ref{VectorisedAV}), that  $\mymat{L}^{\mathrm{v}} = \left(\cancel{\mymat{R}}^{\mathrm{v}}\right)^{\mathrm{t}}$ and in particular 
\begin{equation}
\label{CholeskyIsQRV}
    \mymat{V}\mymat{N}^{\mathrm{v}} = \mymat{Q}^{\mathrm{v}} \left(\mymat{L}^{\mathrm{v}}\right)^{\mathrm{t}}.
\end{equation}
We can perform similar calculations for $\mymat{A}^{\mathrm{m}}$ from Eq.~(\ref{Matrices3}) and the matrix $\mymat{M}$ obtained from the vectors $\{M_{\mathcal{I}}\}$ leading to
\begin{equation}
\label{CholeskyIsQRM}
    \mymat{M}\mymat{N}^{\mathrm{m}} =  \mymat{Q}^{\mathrm{m}} \left(\mymat{L}^{\mathrm{m}}\right)^{\mathrm{t}},\quad \mymat{Q}^{\mathrm{m}}\in\text{O}(D\times s).
\end{equation}

We hence decompose the matrix $\mymat{A}^{\mathrm{vm}}$ from Eq.~(\ref{Matrices2}) as
\begin{equation}
\label{AvmtermDecompose}
    A^{\mathrm{vm}} = \mymat{N}^{\mathrm{v}}\ \mymat{V}^{\mathrm{t}}\ \mymat{M}\ \mymat{N}^{\mathrm{m}} = \mymat{L}^{\mathrm{v}}\left[\left(\mymat{Q}^{\mathrm{v}}\right)^{\mathrm{t}}\mymat{Q}^{\mathrm{m}}\right]\left(\mymat{L}^{\mathrm{m}}\right)^{\mathrm{t}} \equiv \mymat{L}^{\mathrm{v}}\mymat{R}^{\mathrm{vm}}\left(\mymat{L}^{\mathrm{m}}\right)^{\mathrm{t}}.
\end{equation}
The rectangular orthogonal matrices $\mymat{Q}^{\mathrm{v}}$ and $\mymat{Q}^{\mathrm{m}}$ can be parametrized as \citep{doi:10.1021/acs.jpca.5b02015}
\begin{align}
\label{needthis1}
&\mymat{Q}^{\mathrm{v}} = \tilde{\mymat{Q}}^{\mathrm{v}}\mat{1}{D}{d}, \quad \tilde{\mymat{Q}}^{\mathrm{v}}\in\mathrm{O}(D),\\
\label{needthis2}
&\mymat{Q}^{\mathrm{m}} = \tilde{\mymat{Q}}^{\mathrm{m}}\mat{1}{D}{s}, \quad \tilde{\mymat{Q}}^{\mathrm{m}}\in\mathrm{O}(D),
\end{align}
providing the following structure to $\mymat{R}^{\mathrm{vm}}$
\begin{equation}
\label{Rvm}
    \mymat{R}^{\mathrm{vm}} =  \mat{1}{D}{d}^{\mathrm{t}}\ \mymat{Q}^{\mathrm{vm}}\ \mat{1}{D}{s},\quad  \left(\mymat{Q}^{\mathrm{vm}}\right)^{\mathrm{t}}\in\mathrm{O}(D),
\end{equation}
as $\mymat{Q}^{\mathrm{vm}} = \left(\tilde{\mymat{Q}}^{\mathrm{v}}\right)^{\mathrm{t}}\tilde{\mymat{Q}}^{\mathrm{m}}$. 
Finally, by virtue of Eq.~(\ref{AvmtermDecompose}), we can write $\mathcal{Y}(\bm{\Delta},\bm{P})$ as a quadratic form defining the vector  $\bm{B}(\bm{\Delta},\bm{P})$ as
\begin{equation}
        \label{LossQuadratic}
     %\mathcal{Y}(\bm{\Delta};\bm{P}) = \frac{1}{2}\bm{B}^{\mathrm{t}}(\bm{\Delta};\bm{P})\bm{B}(\bm{\Delta};\bm{P}),\\
     \bm{B}(\bm{\Delta},\bm{P}) \equiv \mymat{Q}^{\mathrm{v}}\left(\mymat{L}^{\mathrm{v}}\right)^{\mathrm{t}} \bm{\Delta}^{\mathrm{v}} + \frac{1}{2} \mymat{Q}^{\mathrm{m}}\left(\mymat{L}^{\mathrm{m}}\right)^{\mathrm{t}} \bm{\Delta}^{\mathrm{m}},\quad
     \bm{B}\in\mathbb{R}^{D}.
\end{equation}

We can simplify the expression for $\bm{B}(\bm{\Delta},\bm{P})$ by noticing that $\mathcal{Y}(\bm{\Delta},\bm{P})$ has a symmetry given by the group O($D$)
\begin{align}
\label{SymO}
&\bm{B} \overset{ \text{O}(D)}{\longrightarrow} \mymat{H} \bm{B},\quad \mymat{H}\in\text{O}(D)\\ &\mathcal{Y}_{\mymat{H}} = \left(\mymat{H} \bm{B}\right)^{\mathrm{t}}\left(\mymat{H} \bm{B}\right) = \bm{B}^{\mathrm{t}}\bm{B} =\mathcal{Y}. 
\end{align}
In line with Sec. \ref{FittingFormalism}, we want to reduce the number of actual degrees of freedom (dofs.) by gauging this symmetry via e.g.  $\mymat{Q}^{\mathrm{m}} \equiv \mat{1}{D}{s}$, hence leading to 
\begin{align}
&\mymat{R}^{\mathrm{vm}} =  \mat{1}{D}{d}^{\mathrm{t}}\ \left(\tilde{\mymat{Q}}^{\mathrm{v}}\right)^{\mathrm{t}}\ \mat{1}{D}{s}, \label{Rvmfinal}\\
    \label{Bfinal}
     &\bm{B} = \mymat{Q}^{\mathrm{v}}\left(\mymat{L}^{\mathrm{v}}\right)^{\mathrm{t}} \bm{\Delta}^{\mathrm{v}} + \frac{1}{2}\mat{1}{D}{s}\left(\mymat{L}^{\mathrm{m}}\right)^{\mathrm{t}}\bm{\Delta}^{\mathrm{m}}.
\end{align}
In general, the identification above does not gauge all the degrees of freedom associated to the transformation in Eq.~(\ref{SymO}) as $\mymat{Q}^{\mathrm{m}} \equiv \mat{1}{D}{s}$ fixes  $\text{dim}\left[\text{O}(d\times D)\right] = Ds - s(s+1)/2$ dofs. and $\text{dim}\left[\text{O}(d\times D)\right] \leq \mathrm{dim}\left[\mathrm{O}(D)\right]$ for $s\leq D$. 

At this point, it is not possible to simplify any further our formalism without any information on the  dimension $D = \mathrm{dim}\left[B\right]$.
We set $D \equiv s$ as it minimizes the number of shape parameters to fit for while guaranteeing the positive definiteness of $\mymat{A}^{\mathrm{v}}$ and  $\mymat{A}^{\mathrm{m}}$. Under this assumption, we have that
\begin{align}
&\tilde{\mymat{Q}}^{\mathrm{v}} \in\mathrm{O}(s), \\
&\mymat{R}^{\mathrm{vm}} = \mat{1}{s}{d}^{\mathrm{t}}\ \left(\tilde{\mymat{Q}}^{\mathrm{v}}\right)^{\mathrm{t}}.\label{interesting}
\end{align}
and the gauge $\mymat{Q}^{\mathrm{m}} \equiv \mat{1}{s}{s}$ fixes all the degrees of freedom associated to the symmetry from Eq.~(\ref{SymO}). 

We then parametrize the matrix $\tilde{\mymat{Q}}^{\mathrm{v}}$ as 
\begin{equation}
    \tilde{\mymat{Q}}^{\mathrm{v}}\equiv  \prod_{q=1}^{d}\mymat{H}_{\bm{v}^{(q)}},
\end{equation}
with $\mymat{H}_{\bm{v}^{(q)}}\in\mathbb{R}^{s\times s}$ being the Householder reflector of order $q$. Such operator is determined by a vector $\bm{v}^{(q)}\in\mathbb{R}^{s-q}$ \citep{doi:10.1021/acs.jpca.5b02015}. The components of the vectors $\{\bm{v}^{(q)}\}$ are chosen to be among the shape parameters of our formalism. 

\paragraph*{Loss function and dimensionality of the problem}

We finally summarize our strategy. The variational distribution we designed is 
\begin{align}
    \label{FinaLoss}
    &\mathcal{V}\left(\bm{p}|\bm{P}\right)  \propto \mathrm{exp}\left[-\mathcal{Y}(\bm{\Delta},\bm{P})\right],\\
    &\mathcal{Y}(\bm{\Delta},\bm{P}) = \frac{1}{2}\ \Bigg{|}\ \tilde{\mymat{Q}}^{\mathrm{v}}\mat{1}{s}{d}\left(\mymat{L}^{\mathrm{v}}\right)^{\mathrm{t}} \bm{\Delta}^{\mathrm{v}} + \frac{1}{2}\left(\mymat{L}^{\mathrm{m}}\right)^{\mathrm{t}}\bm{\Delta}^{\mathrm{m}}\ \Bigg{|}^2,\\
    &\bm{\Delta} = \bm{p}-\hat{\bm{p}},\quad
    \tilde{\mymat{Q}}^{\mathrm{v}}\equiv  \prod_{q=1}^{d}\mymat{H}^{(q)}.
\end{align}
and it is parametrized in terms of the following vector of shape parameters 
\begin{multline} 
\label{VecP}
    \bm{P} = \Big{\{}
    \big{\{}\bm{v}_{\alpha}\ \big{\}}_{\alpha=1,\dots,d-1},
    \big{\{}\ \bm{m}_{\mathcal{I}}\ \big{\}}_{\mathcal{I}=1,\dots,s-1},
    \big{\{}\bm{v}^{(q)}\big{\}}_{q=1,\dots,d},\\
    \big{\{}\ V_{\alpha}\ \big{\}}_{\alpha = 1,\dots,d},  \{ M_{\mathcal{I}}\ \big{\}}_{\mathcal{I} = 1,\dots,s},
    \hat{\bm{p}}
    \Big{\}}.
\end{multline}
In total, for a parameter space of dimension $d=\mathrm{dim}\left[\bm{p}\right]$, the associated number of variational parameters is 
\begin{equation}
\label{Ndof}
    N_{\mathrm{shape}} \equiv \text{dim}\left[\bm{P}\right] = \frac{1}{8}\left(10d + 7d^2 + 6d^3 + d^4\right).
\end{equation}
In App.~\ref{DOFsDescription}, we summarize how different parts of the variational distribution from Eq.~(\ref{FinaLoss}) depend on the shape parameters.
Starting from a set of randomly sampled points,
\begin{equation}
    \Big{\{}\ \mathcal{P}_i;\bm{p}_i\Big{\}}_{i=1,\dots,n_{\mathrm{s}}},\quad \mathcal{P}_i = \mathcal{P}\left(\bm{p}_{i}|\bm{x}\right)\label{SampleSet},
\end{equation}
we identify the optimal shape parameters $\bm{P}$ such that ${\mathcal{V}\left(\bm{p}|\bm{P}\right) \approx \mathcal{P}\left(\bm{p}\right)}$
by minimizing a quadratic loss function,
\begin{equation}
\label{LossFunction}
    \Phi\left(\bm{P}\right) = \frac{1}{n_{\mathrm{s}}} \sum_{i=1}^{n_{\mathrm{s}}} \phi^2\left(\mathcal{P}_i,\mathcal{V}_i\right),\quad \mathcal{V}_i \equiv \mathcal{V}\left(\bm{p}_i|\bm{P}\right).
\end{equation}
Depending on the application, different definitions can be chosen for the scalar function $\phi$, which we call cost. In this paper, we choose
\begin{equation}
    \label{our_cost}
    \phi\left(\mathcal{P}_i,\mathcal{V}_i\right) = \left(\frac{| \mathcal{V}_i-\mathcal{P}_i|}{| \mathcal{V}_i|+|\mathcal{P}_i|}\right)^{\frac{1}{2}}.
\end{equation}

\section{(Re-)construction of the KiDS-450 posterior}
\label{TestingMethod}
\begin{figure*}
\centering
\vspace{-0.cm}
\begin{tabular}{c}
\multicolumn{1}{c}{\hspace{.6cm}\includegraphics[width=0.77\textwidth,trim= 0 0 0 0.1cm,clip]{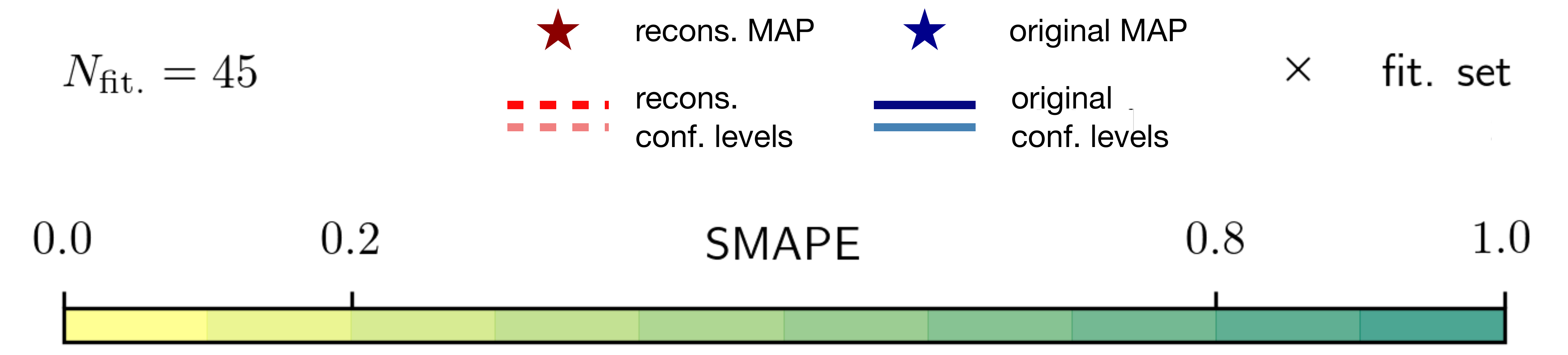}}\vspace{-0.cm}\\
\vspace{-0.cm}\hspace{-0.5cm}\includegraphics[width=0.34\textwidth,trim= 0 0 0 0,clip]{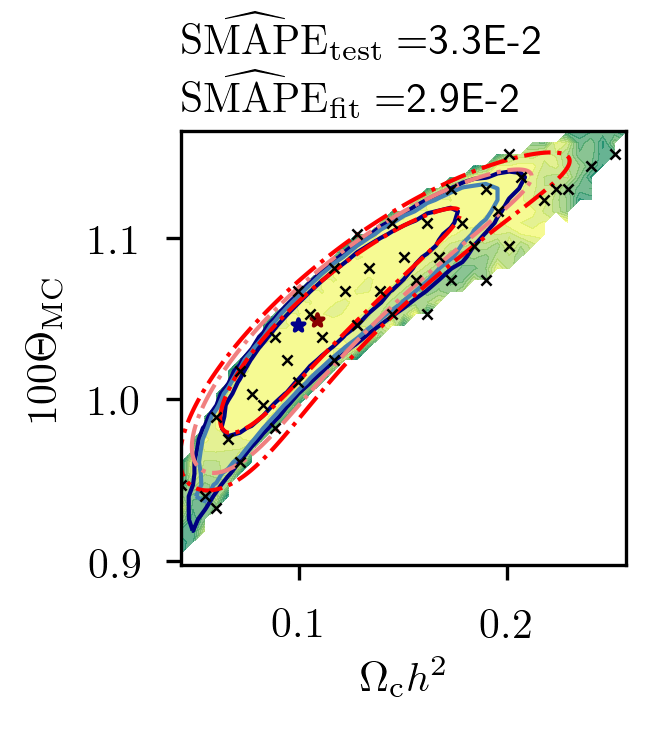}\hspace{-0.2cm}\includegraphics[width=0.34\textwidth,trim= 0cm 0 0 0cm,clip]{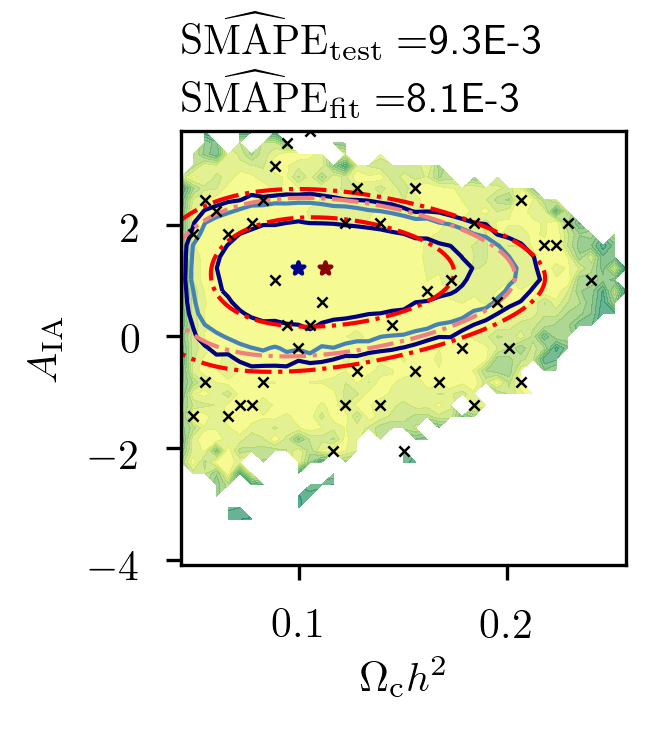}\hspace{-0.2cm}\includegraphics[width=0.34\textwidth,trim= 0 0 0 0cm,clip]{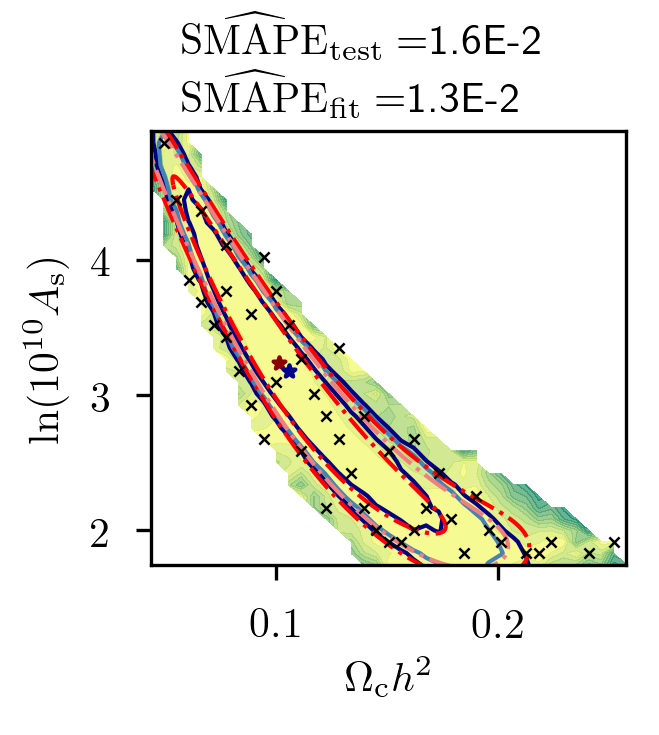}\vspace{-0cm}\\
\vspace{-0.35cm}\hspace{-0.5cm}\includegraphics[width=0.34\textwidth,trim= 0 0 0 0cm,clip]{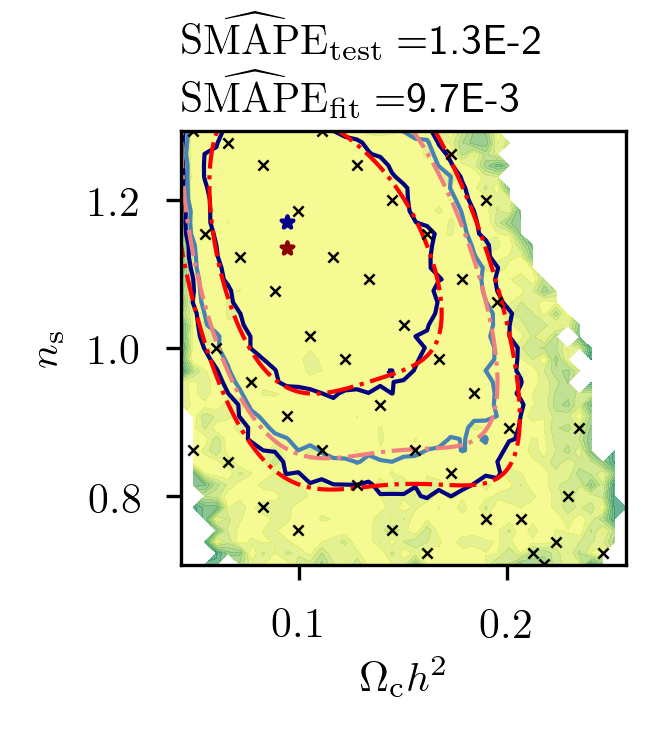}\hspace{-0.2cm}\includegraphics[width=0.34\textwidth,trim= 0cm 0 0 0cm,clip]{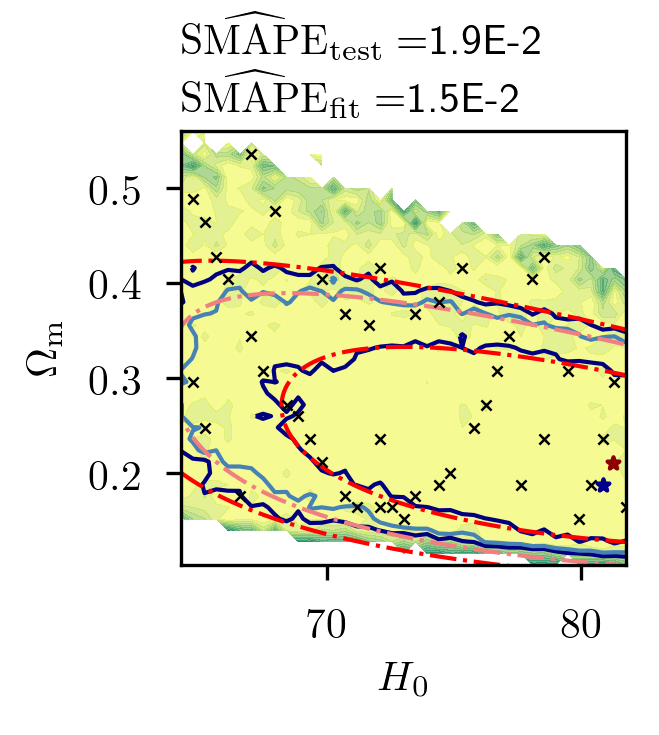}\hspace{-0.2cm}\includegraphics[width=0.34\textwidth,trim= 0 0 0 0cm,clip]{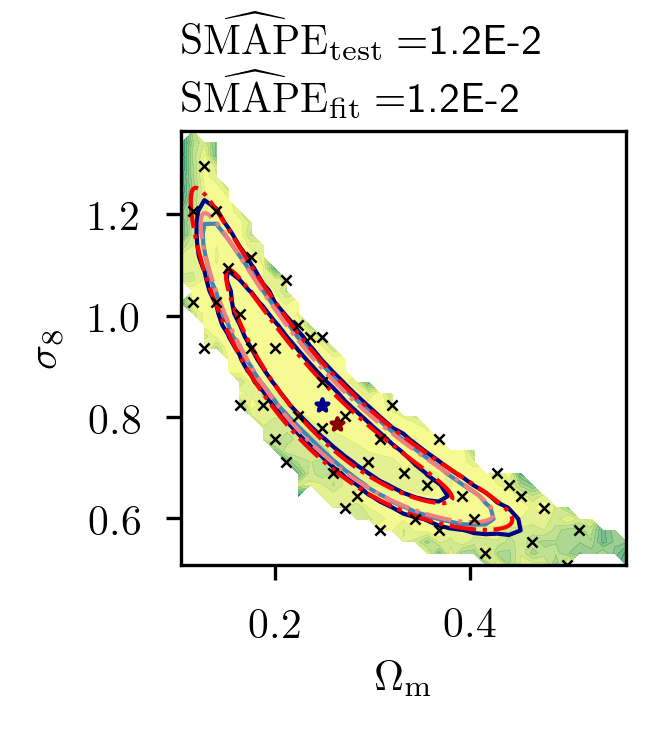}\vspace{-0cm}\\
\end{tabular}
\caption{Constructing two-dimensional KiDS-450 marginals from $N_{\rm fit} = 45$ posterior samples only. The points used for construction are indicated by black crosses, while the 68\%, 90\% and 95\% confidence levels obtained via the variational-Bayes posterior are depicted in shades of red. The confidence levels referring the original KiDS-450 posterior, are depicted in shades of blue, these being obtained  by marginalising the original MCMC chain of $\sim3\times10^6$ (accepted and rejected) samples. The dark red and blue stars respectively indicate the MAP for the reconstructed and the original posterior. The yellow and green background map indicates that differences between the two approaches only affect the posterior tails. 
}
\label{Fig1}
\end{figure*}

\begin{figure*}
\centering
\vspace{-0.cm}
\begin{tabular}{c}
\multicolumn{1}{c}{\hspace{.6cm}\includegraphics[width=0.77\textwidth,trim= 0 0 0 0.1cm,clip]{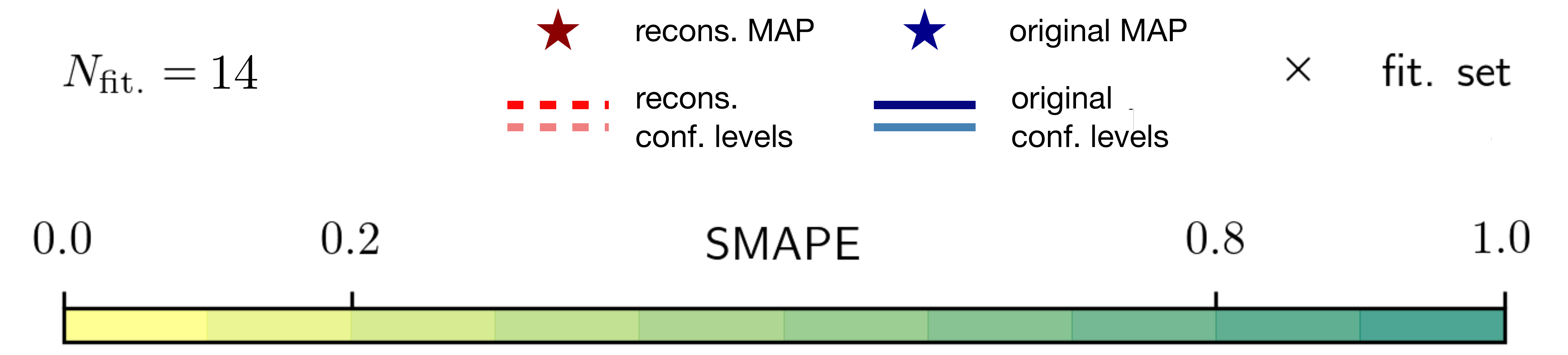}}\vspace{-0.cm}\\
\vspace{-0.cm}\hspace{-0.5cm}\includegraphics[width=0.34\textwidth,trim= 0 0 0 0cm,clip]{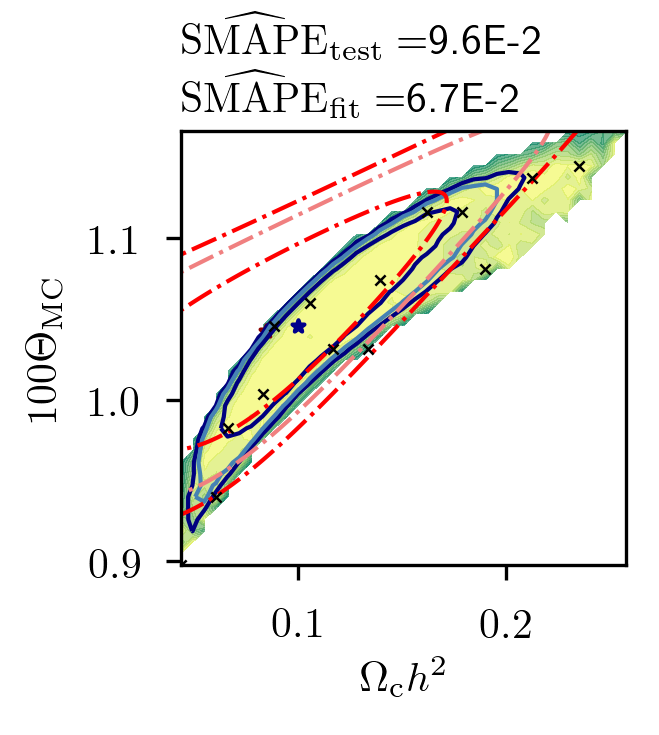}\hspace{-0.2cm}\includegraphics[width=0.34\textwidth,trim= 0cm 0 0 0cm,clip]{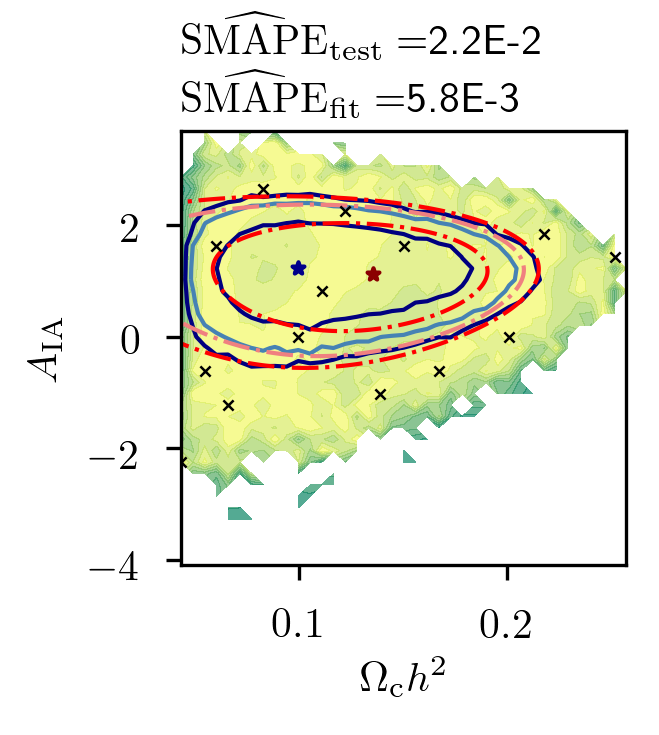}\hspace{-0.2cm}\includegraphics[width=0.34\textwidth,trim= 0 0 0 0cm,clip]{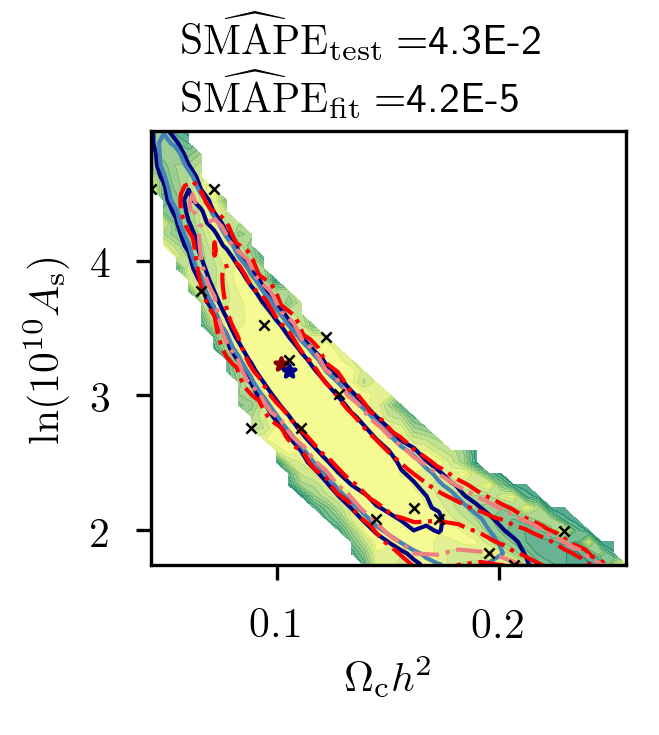}\vspace{-0cm}\\
\vspace{-0.35cm}\hspace{-0.5cm}\includegraphics[width=0.34\textwidth,trim= 0 0 0 0cm,clip]{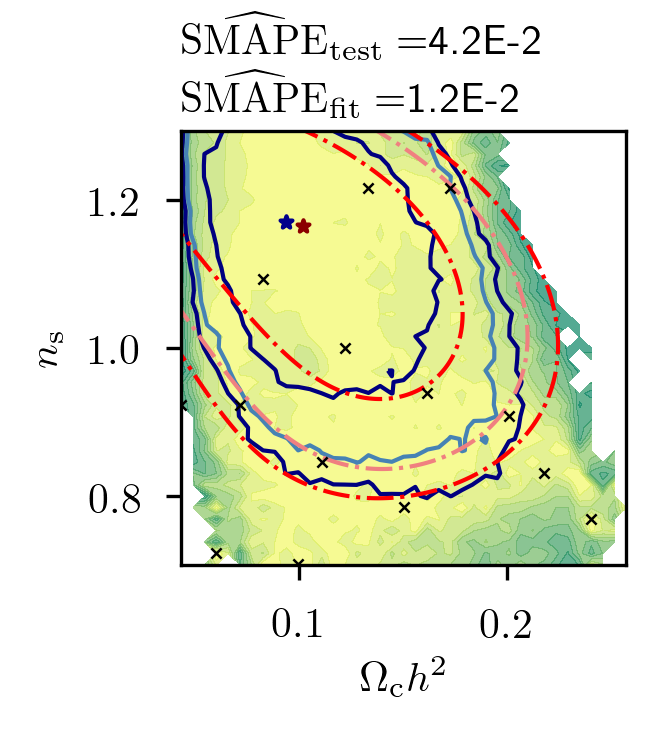}\hspace{-0.2cm}\includegraphics[width=0.34\textwidth,trim= 0cm 0 0 0cm,clip]{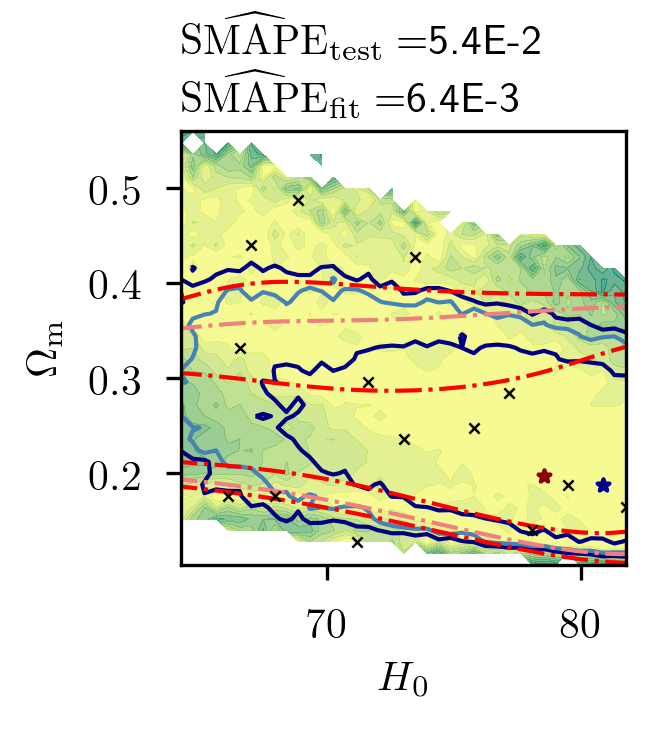}\hspace{-0.2cm}\includegraphics[width=0.34\textwidth,trim= 0 0 0 0cm,clip]{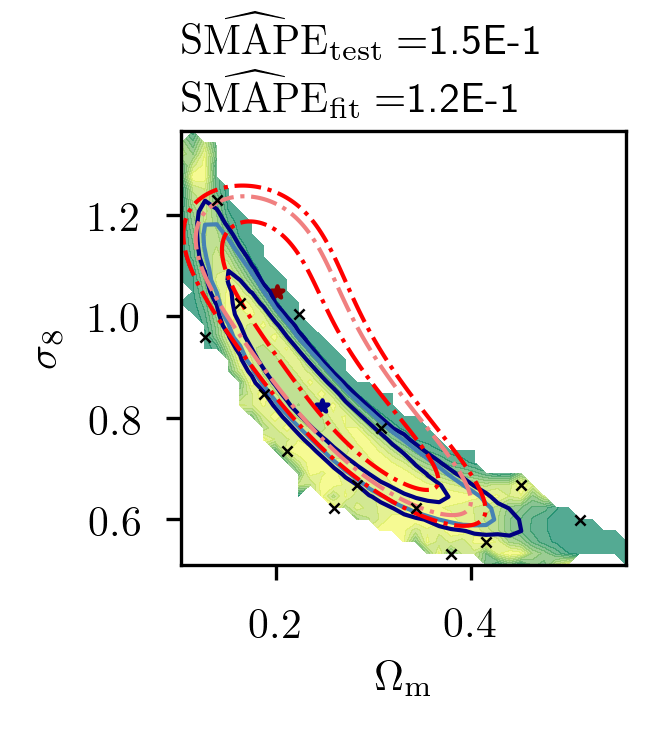}\vspace{-0cm}\\
\end{tabular}
\caption{Stress-testing the variational-Bayes solution by providing the absolute minimum number of posterior samples: the variational distribution has 14 free parameters, hence we provided only $N_{\rm fit} = 14$ points. Comparing the original confidence levels (shades of blue) to the ensuing variational-Bayes ones (shades of red), we find a mixture of good and insufficient reconstructions of the true posterior shape. This illustrates how rapidly the variational distribution converges towards the true posterior. The dark red and blue stars respectively indicate the MAP for the reconstructed and the original posterior.}
\label{Fig2}
\end{figure*}

In this section, we apply our variational-Bayes approach on the  KiDS-450 analysis of \citet{2017MNRAS.465.1454H}, which provides a public MCMC chain\footnote{\label{note2}\url{http://kids.strw.leidenuniv.nl/cs2016/MCMC_README.html}} consisting of $\sim4\times10^5$ accepted samples (after burn-in removal) from a 7-dimensional posterior. When accounting also for rejected points and including the burn-in, the provided chain counts $\sim3\times10^6$ samples, which represents the true numerical cost of this MCMC run.

We split the samples in a test set of $N_{\mathrm{test}}$ samples and a smaller fitting set of $N_{\mathrm{fit}}$ samples. The variational-Bayes algorithm fits the variational distribution only to the fitting set. The test set would in a realistic application not be available and is here used only to evaluate the deviation between our variational-Bayes posterior and the KiDS-450 MCMC posterior.

We compare the two posteriors by a symmetrized version of a percentual deviation, known as SMAPE (Symmetrized Mean Absolute Percentage Error).
At each point $\bm{p}_i$ of the test set, the final variational-Bayes posterior $\mathcal{V}(\bm{p}_i|\bm{x}, \hat{\bm{P}})$ is compared to the MCMC posterior $\mathcal{P}(\bm{p}_i|\bm{x})$ by the SMAPE, defined as
\begin{equation}
    \label{SMAPE}
    \mathrm{SMAPE}_i \equiv \frac{| \mathcal{V}(\bm{p}_i|\bm{x}, \hat{\bm{P}})-\mathcal{P}(\bm{p}_i|\bm{x})|}{| \mathcal{V}(\bm{p}_i|\bm{x}, \hat{\bm{P}})|+|\mathcal{P}(\bm{p}_i|\bm{x})|}.
\end{equation}
This SMAPE can take values between 0 (exact agreement) and 1 (maximally discrepant). It is similar to a percentual deviation, only that it correctly reports a non-zero deviation also when either of the two compared quantities can take zero values. The SMAPE will therefore detect it when either of the two posteriors vanishes while the other does not.

The local SMAPE at each point can then be averaged into a single global estimate:
\begin{equation}
     \label{averagedSMAPE}
     \reallywidehat{\mathrm{SMAPE}} = \frac{1}{N}\sum_{i=1}^N  \mathrm{SMAPE}_i\ w_i,
\end{equation}
where we introduced the weight
\begin{equation}
     w_i = \mathrm{max}\left(\mathcal{P}(\bm{p}_i|\bm{x}),\mathcal{V}(\bm{p}_i|\bm{x}, \hat{\bm{P}})\right). 
\end{equation}
The introduction of this weight causes points of high posterior value to dominate the global SMAPE estimate. Simultaneously, points in low density regions have a lower impact. In this way, the global SMAPE correctly reports the quality of posterior reconstruction within a few standard deviations from the peak, rather than reporting the quality of reconstruction in the posterior tails.

\subsection{Two-dimensional posteriors}

We begin with the simpler task of reconstructing six 2D marginals derived from the otherwise 7-dimensional KiDS-450 posterior. The MCMC marginals were obtained by binning the KiDS chain. A few points were then randomly selected and provided to the variational-Bayes algorithm to fit.

The fitter updates the values of the shape parameters $\boldsymbol{P}$ and requires a loss function  $\phi$ to be minimized. We chose Eq.~(\ref{averagedSMAPE}) as the loss function, noting that this choice can be exchanged. 

Figs.~\ref{Fig1} and \ref{Fig2} display the original MCMC posterior in blue, and our variational-Bayes posterior in red. In Fig.~\ref{Fig1}, we considered  ${N_{\mathrm{fit}} \approx 3 N_{\mathrm{shape}}}$ (45 fitting points). In Fig.~\ref{Fig2}, we tested the variational-Bayes approach under the extreme condition of providing the minimally possible number of fitting points, namely $N_{\mathrm{fit}} = N_{\mathrm{shape}} = 14$.

In both these cases, the number of posterior evaluations is orders of magnitude smaller than for a typical MCMC chain. We mark the two posterior peaks with a red or blue star and indicate the 68\%, 90\% and 95\% contours.  The position of the fitting points is indicated by black crosses. The green background colour map reflects the local value for the SMAPE from Eq.~(\ref{SMAPE}). 

In Fig.~\ref{Fig1}, we employ (roughly) three times more fitting points than shape parameters to be optimized. In all studied cases, this led to the variational-Bayes solution being virtually indistinguishable from the original MCMC posterior. 

As the accuracy of the variational-Bayes solution must decrease for decreasing number of fitting points, we study in Fig.~\ref{Fig2} the limiting case of  $N_{\mathrm{fit}} = N_{\mathrm{shape}}$. As expected, the quality of the variational-Bayes solution then becomes highly case dependent: it may or may not be an accurate representation of the true posterior. 

While the purpose of the current work is to formally obtain the expression in Eq.~\eqref{FinaLoss}  and prove its applicability to realistic use cases, we defer to future work a thorough study on the robustness of the reconstruction accuracy against the minimum number of training samples and their distribution.

\subsection{7-dimensional posterior}

\begin{figure*}
\centering
\vspace{-0.cm}
\begin{tabular}{c}
\multicolumn{1}{c}{\includegraphics[width=\textwidth]{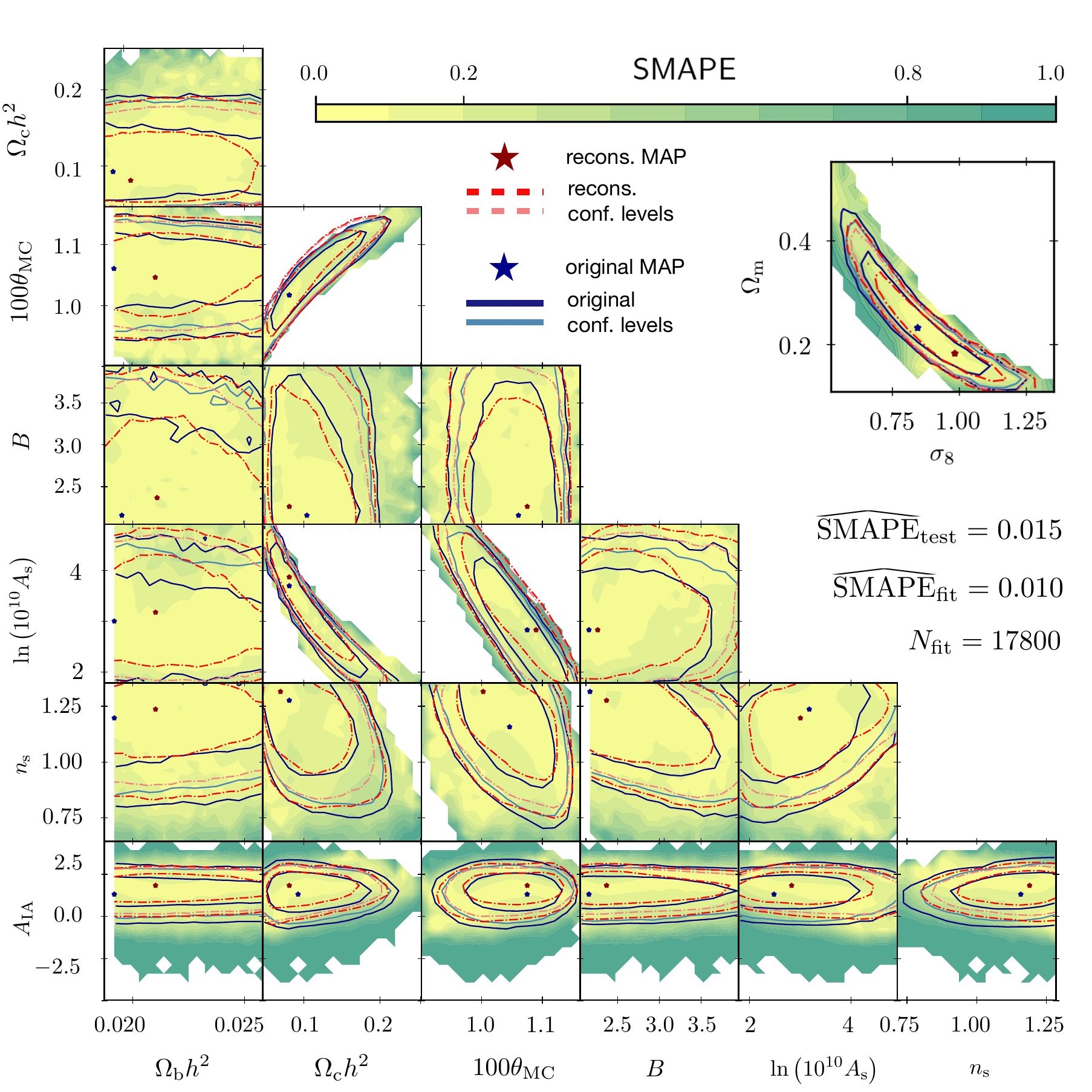}}
\end{tabular}
\caption{\textit{Bottom-left:} Corner plot showing the two-dimensional marginals for all the primary parameters considered in the original KiDS-450 tomographic weak gravitational lensing analysis. \textit{Top-right:} marginal distribution for the  derived parameters $\{\sigma_8,\Omega_{\mathrm{m}}\}$. In each subplot, we compare the marginal distribution from the full MCMC in light blue (solid line) and our variational-Bayes solution in red (dot-dashed line). The variational-Bayes solution approximates the full 7-dimensional posterior, although only its marginals are visible here.  The number of shape parameters is $N_{\mathrm{shape}} \approx 600$. The number of posterior samples employed for fitting is $N_{\mathrm{fit}}= 17800$. We also specified the posterior to be zero at 8900 further positions in order to deal with the curse of dimensionality. This means that we set the posterior to zero in the vast and empty corners of the 7-dimensional space. The variational-Bayes solution accordingly only required $0.6\%$ of the MCMC's computational cost.
}
\label{Fig4}
\end{figure*}

In this section, we apply the variational-Bayes approach to the construction of the joint 7-dimensional posterior of the  KiDS-450 analysis \citep{2017MNRAS.465.1454H}. According to Eq.~(\ref{Ndof}), the number of shape parameters to fit for is $N_{\mathrm{shape}} = 608$.

In higher dimensions, we found that the speed of the fitting procedure can be vastly accelerated if the fitter is aided in the following two points.

Firstly, the slowest shape parameters to estimate are the coordinates of the posterior peak $\hat{\boldsymbol{p}}$. As $\hat{\boldsymbol{p}}$ enters the variational distribution through $\boldsymbol{\Delta}= \boldsymbol{p}-\hat{\boldsymbol{p}}$, each time the fitter updates the coordinates of the peak, all other shape parameters lead to a momentarily worse fit and have to be updated too. As there are many more shape parameters other than the peak coordinates, this leads to a slow-down of the fitter which can be prevented if the peak coordinates are estimated first. When providing the coordinates of the peak known from the MCMC chain, we therefore observed a speed-up of the fitter.

Secondly, the fitter's speed can be accelerated when providing it with extra information in order to alleviate the burden resulting from the curse of dimensionality. The curse of dimensionality describes that the corners of an $n$-dimensional space take up ever larger fractions in comparison to the volume of an $n$-sphere. In higher dimensions, the fitter will thus encounter ever larger volumes about which it has no information. More fitting points must then be provided.
However, we here argue that those additional fitting points do not need to stem from an evaluation of the likelihood; the remote corners of the $n$-dimensional space have negligible probability anyhow, making an evaluation of the expensive likelihood unnecessary. To assist the fitter in handling the curse of dimensionality, we thus provide it with artificial points where we set the posterior density to zero.

The triangular plot in Fig. \ref{Fig4} shows the result offered by the variational-Bayes approach for the 7-dimensional KiDS-450 posterior. We employed $N_{\mathrm{fit}} = 17800$ posterior points randomly selected from the KiDS-450 MCMC chain. These are points where the actual likelihood code was run.  We augment these by 8900 points where we set the posterior density to zero. This setup allows us to reach a final score of $\reallywidehat{\mathrm{SMAPE}}_{\mathrm{test}} = 0.015$ on the selected samples. The number of likelihood evaluations used in our model is $0.6\%$ of the total original MCMC posterior samples.

\begin{figure*}
\centering
\vspace{-0.cm}
\begin{tabular}{c}
\includegraphics[width=1\textwidth,trim= 0 0 0 0cm,clip]{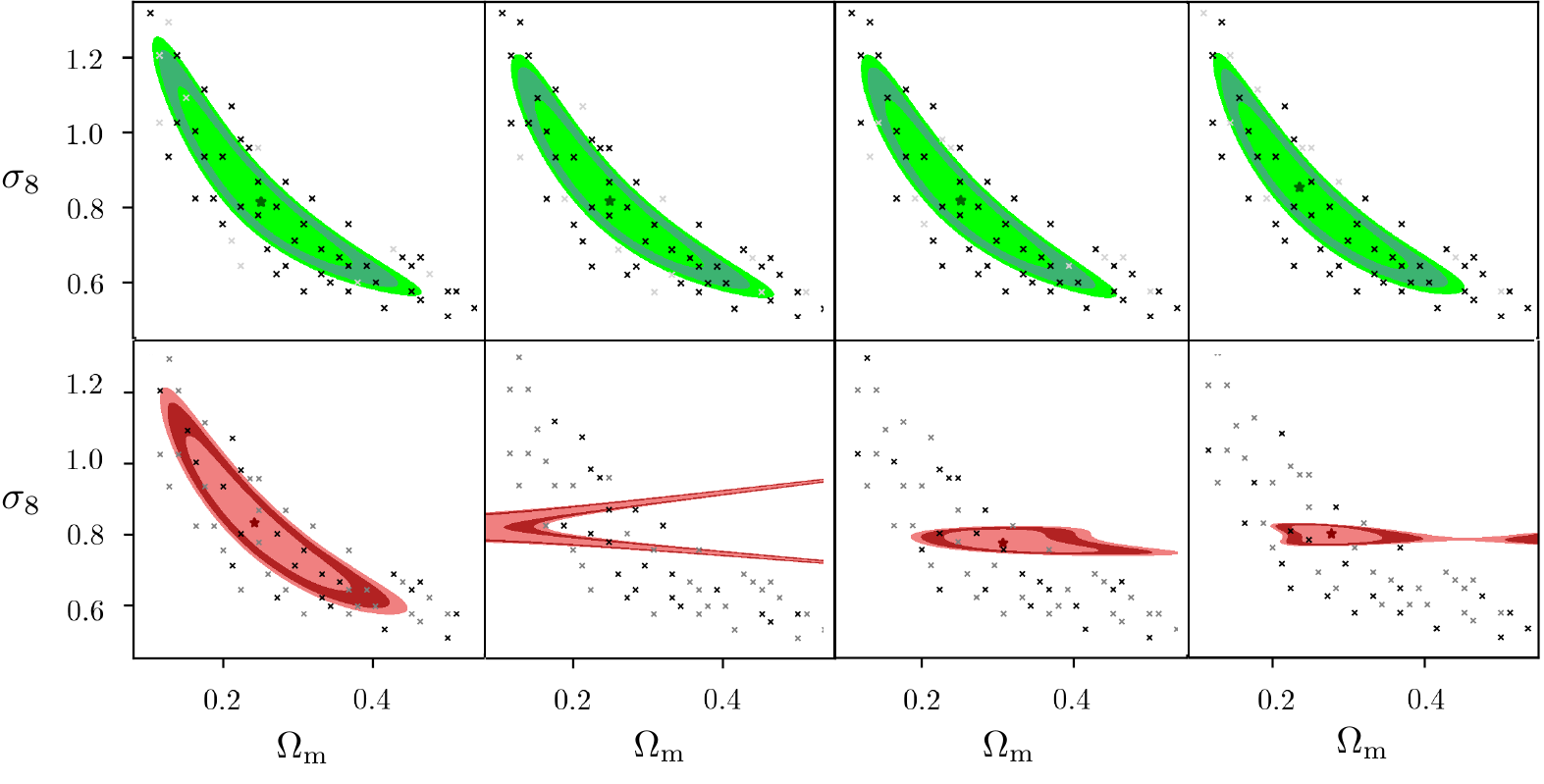}
\end{tabular}
\caption{Illustration of the Leave-N-out test to judge the quality of the  variational-Bayes posterior. Out of an initial set of 55 points (black crosses), each panel of the first row selected randomly $N_{\mathrm{fit}}=45$ points, with the discarded 10 points being greyed out. As the posterior shape barely varies between the panels, using all 55 points for a joint reconstruction will yield a high-quality posterior. In the lower row, only 22 points were randomly chosen. The Leave-N-out test then indicates that these were too few points to trust the posterior reconstruction, as each panel displays a differently shaped posterior.}
\label{Fig:leave_n_out}
\end{figure*}

\section{Discussion and conclusions}
\label{discANDconcl}
\subsection{Discussions}
\label{disc}
We have formally derived a novel variational-Bayes solution to compute non-Gaussian posteriors from extremely
expensive likelihoods. We explored its geometrical properties and applied it to a realistic use case. We made a first attempt in evaluating the reconstruction performance of the method proposed optimising the parameters of the model against sub-samples of the original MCMC chains. In this final section we pave the way to future studies aiming at validating the method via suited performance metrics and statistical tests.

In the figures presented as support to our analysis, it 
is easy to judge the quality of the variational-Bayes posterior, as it can be compared to the original MCMC posterior. In a realistic setup, the MCMC posterior would not be available for comparison, and the quality of the variational-Bayes posterior has to be judged from the few likelihood evaluations of the fitting set only.
One summary statistic that can be evaluated from the fitting set without comparison to an MCMC run is the global SMAPE from Eq.~(\ref{averagedSMAPE}), which we indicate in our figures. We found that values below 0.02 can indicate an accurate posterior reconstruction, but counter examples exist. For example, Fig.~\ref{Fig2} evidences that global SMAPE values of $10^{-4}$ can still yield inaccurate multimodal posteriors. 
In general, performance metrics evaluated on the fitting set do not assess how well the model generalizes out-of-sample. 
Accordingly, we found the most reliable intrinsic judgement of fitting quality to be a Leave-N-Out approach. We divide the fitting set into $N$ random subsets with fixed cardinality. We then repeat the variational-Bayes optimization multiple times, randomly leaving out one of the $N$ subsets, being subsequently employed as test set for performance evaluation. If all fits yield the same posterior, then the entire fitting set is likely to include sufficiently many samples to guarantee a successful variational-Bayes posterior. In contrast, if leaving points out yields a drastically different posterior, then insufficiently many likelihood evaluations are available. If possible, more samples should then be added in the area where differences in the estimated posteriors are seen. In Fig.~\ref{Fig:leave_n_out}, we employ the Leave-N-out verification to validate the reconstruction of the marginal posterior $\{\sigma_8,\Omega_{\mathrm{m}}\}$ in Fig.~\ref{Fig1}. For this test, 10 random test sets of 10 points were obtained from an initial set of 55 points. 45 being points sufficient for the optimization, we obtained 10 extremely similar reconstructions, four of which are proposed in Fig.~\ref{Fig:leave_n_out}.

Throughout our work, we relied on the global SMAPE as metric for evaluating the divergence between reconstructed and original posterior. This is a general purpose metric and for future work we advise to relate it to relevant summary statistics such as the mean and standard deviation of the projected marginal distributions. Furthermore, follow up studies should aim at generalising the preliminary analysis carried out in this section to identify the minimum number of samples which ensures satisfactory reconstruction performance regardless the distribution of the fitting points.

\subsection{Conclusions}
\label{concl}
We have developed a variational-Bayes solution building upon the work of \citetalias{DALI} and \citetalias{DALII}, with the aim of constructing posteriors when Monte Carlo Markov Chains cannot be run, either due to run times (extremely expensive likelihoods), or because the physical model can only be evaluated at discrete locations in parameter space (simulation-based inference). 

A variational-Bayes approach fits a so-called `variational distribution' to a set of posterior samples, thereby determining suitable values for the tunable shape-parameters of the variational distribution. The best-fitting variational distribution is taken as representation of the posterior.

We applied our algorithm on the KiDS-450 analysis \citep{2017MNRAS.465.1454H} which originally used $\sim3\times10^6$ MCMC (accepted and rejected) samples (including burn-in) to construct their posteriors.

In comparison, we constructed accurate 2D marginals of the KiDS posterior from 14 to 45 samples, and the joint 7-dimensional posterior from $\approx 18000$ samples.

A future aim is to minimize the number of required samples even further. 

\section*{Acknowledgements}
% Entry for the table of contents, for this guide only
\addcontentsline{toc}{section}{Acknowledgements}
M. R. thanks Guilhem Lavaux for support with the GSL.  We thank Simon Portegies Zwart for fruitful discussions and sharing of numerical infrastructure. We convey our gratitude to Doogesh Kodi Ramanah for constructive suggestions to improve the manuscript. This work has made use of the Horizon Cluster hosted by Institut d’Astrophysique de Paris. We thank Stephane Rouberol for maintenance and running of this computing cluster.

\section*{Data availability}
\addcontentsline{toc}{section}{Data availability}
The data underlying this article are available at \url{http://kids.strw.leidenuniv.nl/cs2016/MCMC_README.html}

%%%%%%%%%%%%%%%%%%%% REFERENCES %%%%%%%%%%%%%%%%%%
\bibliographystyle{mnras}
\bibliography{example}

\begin{thebibliography}{}
\makeatletter
\relax
\def\mn@urlcharsother{\let\do\@makeother \do\$\do\&\do\#\do\^\do\_\do\%\do\~}
\def\mn@doi{\begingroup\mn@urlcharsother \@ifnextchar [ {\mn@doi@}
  {\mn@doi@[]}}
\def\mn@doi@[#1]#2{\def\@tempa{#1}\ifx\@tempa\@empty \href
  {http://dx.doi.org/#2} {doi:#2}\else \href {http://dx.doi.org/#2} {#1}\fi
  \endgroup}
\def\mn@eprint#1#2{\mn@eprint@#1:#2::\@nil}
\def\mn@eprint@arXiv#1{\href {http://arxiv.org/abs/#1} {{\tt arXiv:#1}}}
\def\mn@eprint@dblp#1{\href {http://dblp.uni-trier.de/rec/bibtex/#1.xml}
  {dblp:#1}}
\def\mn@eprint@#1:#2:#3:#4\@nil{\def\@tempa {#1}\def\@tempb {#2}\def\@tempc
  {#3}\ifx \@tempc \@empty \let \@tempc \@tempb \let \@tempb \@tempa \fi \ifx
  \@tempb \@empty \def\@tempb {arXiv}\fi \@ifundefined
  {mn@eprint@\@tempb}{\@tempb:\@tempc}{\expandafter \expandafter \csname
  mn@eprint@\@tempb\endcsname \expandafter{\@tempc}}}

\bibitem[\protect\citeauthoryear{{Abbott} et~al.,}{{Abbott}
  et~al.}{2016}]{DESnoses}
{Abbott} T.,  et~al., 2016, \mn@doi [\prd] {10.1103/PhysRevD.94.022001}, \href
  {https://ui.adsabs.harvard.edu/abs/2016PhRvD..94b2001A} {94, 022001}

\bibitem[\protect\citeauthoryear{{Amendola} \& {G{\'o}mez-Valent}}{{Amendola}
  \& {G{\'o}mez-Valent}}{2020}]{Luca}
{Amendola} L.,  {G{\'o}mez-Valent} A.,  2020, \mn@doi [\mnras]
  {10.1093/mnras/staa2362}, \href
  {https://ui.adsabs.harvard.edu/abs/2020MNRAS.498..181A} {498, 181}

\bibitem[\protect\citeauthoryear{{Audren}, {Lesgourgues}, {Bird}, {Haehnelt}
  \& {Viel}}{{Audren} et~al.}{2013}]{JulienForecastNeutrinos}
{Audren} B.,  {Lesgourgues} J.,  {Bird} S.,  {Haehnelt} M.~G.,   {Viel} M.,
  2013, \mn@doi [\jcap] {10.1088/1475-7516/2013/01/026}, \href
  {https://ui.adsabs.harvard.edu/abs/2013JCAP...01..026A} {2013, 026}

\bibitem[\protect\citeauthoryear{{Durrer} \& {Maartens}}{{Durrer} \&
  {Maartens}}{2008}]{RuthRoy}
{Durrer} R.,  {Maartens} R.,  2008, \mn@doi [General Relativity and
  Gravitation] {10.1007/s10714-007-0549-5}, \href
  {https://ui.adsabs.harvard.edu/abs/2008GReGr..40..301D} {40, 301}

\bibitem[\protect\citeauthoryear{{Hagstotz}, {de Salas}, {Gariazzo}, {Gerbino},
  {Lattanzi}, {Vagnozzi}, {Freese}  \& {Pastor}}{{Hagstotz}
  et~al.}{2020}]{Steffen}
{Hagstotz} S.,  {de Salas} P.~F.,  {Gariazzo} S.,  {Gerbino} M.,  {Lattanzi}
  M.,  {Vagnozzi} S.,  {Freese} K.,   {Pastor} S.,  2020, arXiv e-prints, \href
  {https://ui.adsabs.harvard.edu/abs/2020arXiv200302289H} {p. arXiv:2003.02289}

\bibitem[\protect\citeauthoryear{{Hildebrandt} et~al.,}{{Hildebrandt}
  et~al.}{2017}]{2017MNRAS.465.1454H}
{Hildebrandt} H.,  et~al., 2017, \mn@doi [\mnras] {10.1093/mnras/stw2805},
  \href {https://ui.adsabs.harvard.edu/abs/2017MNRAS.465.1454H} {465, 1454}

\bibitem[\protect\citeauthoryear{{LSST Science Collaboration} et~al.,}{{LSST
  Science Collaboration} et~al.}{2009}]{2009arXiv0912.0201L}
{LSST Science Collaboration} et~al., 2009, preprint, \href
  {http://adsabs.harvard.edu/abs/2009arXiv0912.0201L} {} (\mn@eprint {arXiv}
  {0912.0201})

\bibitem[\protect\citeauthoryear{{Laureijs} et~al.,}{{Laureijs}
  et~al.}{2011}]{2011arXiv1110.3193L}
{Laureijs} R.,  et~al., 2011, preprint, \href
  {http://adsabs.harvard.edu/abs/2011arXiv1110.3193L} {} (\mn@eprint {arXiv}
  {1110.3193})

\bibitem[\protect\citeauthoryear{MacKay}{MacKay}{2002}]{10.5555/971143}
MacKay D. J.~C.,  2002, Information Theory, Inference \& Learning Algorithms.
Cambridge University Press, USA

\bibitem[\protect\citeauthoryear{{Mandelbaum}, {Miyatake}, {Hamana}, {Oguri},
  {Simet}  \& {Utsumi}}{{Mandelbaum} et~al.}{2017}]{HSC}
{Mandelbaum} R.,  {Miyatake} H.,  {Hamana} T.,  {Oguri} M.,  {Simet} M.,
  {Utsumi} Y.,  2017, preprint (\mn@eprint {arXiv} {1705.06745})

\bibitem[\protect\citeauthoryear{{Sellentin}}{{Sellentin}}{2015}]{DALII}
{Sellentin} E.,  2015, \mn@doi [MNRAS] {10.1093/mnras/stv1671}, \href
  {http://adsabs.harvard.edu/abs/2015MNRAS.453..893S} {453, 893}

\bibitem[\protect\citeauthoryear{{Sellentin} \& {Durrer}}{{Sellentin} \&
  {Durrer}}{2015}]{RuthNeutrinos}
{Sellentin} E.,  {Durrer} R.,  2015, \mn@doi [\prd]
  {10.1103/PhysRevD.92.063012}, \href
  {https://ui.adsabs.harvard.edu/abs/2015PhRvD..92f3012S} {92, 063012}

\bibitem[\protect\citeauthoryear{{Sellentin} \& {Sch{\"a}fer}}{{Sellentin} \&
  {Sch{\"a}fer}}{2016}]{WBjoern}
{Sellentin} E.,  {Sch{\"a}fer} B.~M.,  2016, \mn@doi [\mnras]
  {10.1093/mnras/stv2805}, \href
  {https://ui.adsabs.harvard.edu/abs/2016MNRAS.456.1645S} {456, 1645}

\bibitem[\protect\citeauthoryear{{Sellentin}, {Quartin}  \&
  {Amendola}}{{Sellentin} et~al.}{2014}]{DALI}
{Sellentin} E.,  {Quartin} M.,   {Amendola} L.,  2014, \mn@doi [MNRAS]
  {10.1093/mnras/stu689}, \href
  {http://adsabs.harvard.edu/abs/2014MNRAS.441.1831S} {441, 1831}

\bibitem[\protect\citeauthoryear{Shepard, Brozell  \& Gidofalvi}{Shepard
  et~al.}{2015}]{doi:10.1021/acs.jpca.5b02015}
Shepard R.,  Brozell S.~R.,   Gidofalvi G.,  2015, \mn@doi [The Journal of
  Physical Chemistry A] {10.1021/acs.jpca.5b02015}, 119, 7924

\bibitem[\protect\citeauthoryear{Tegmark, Taylor  \& Heavens}{Tegmark
  et~al.}{1997}]{Tegmark:1996bz}
Tegmark M.,  Taylor A.,   Heavens A.,  1997, \mn@doi [Astrophys.J.]
  {10.1086/303939}, 480, 22

\makeatother
\end{thebibliography}

%%%%%%%%%%%%%%%%%%%% App. %%%%%%%%%%%%%%%%%%
\appendix
\section{Number of shape parameters}
\label{DOFsDescription}
\begin{figure}
\label{DimScaling}
\centering
\includegraphics[width=0.4\textwidth]{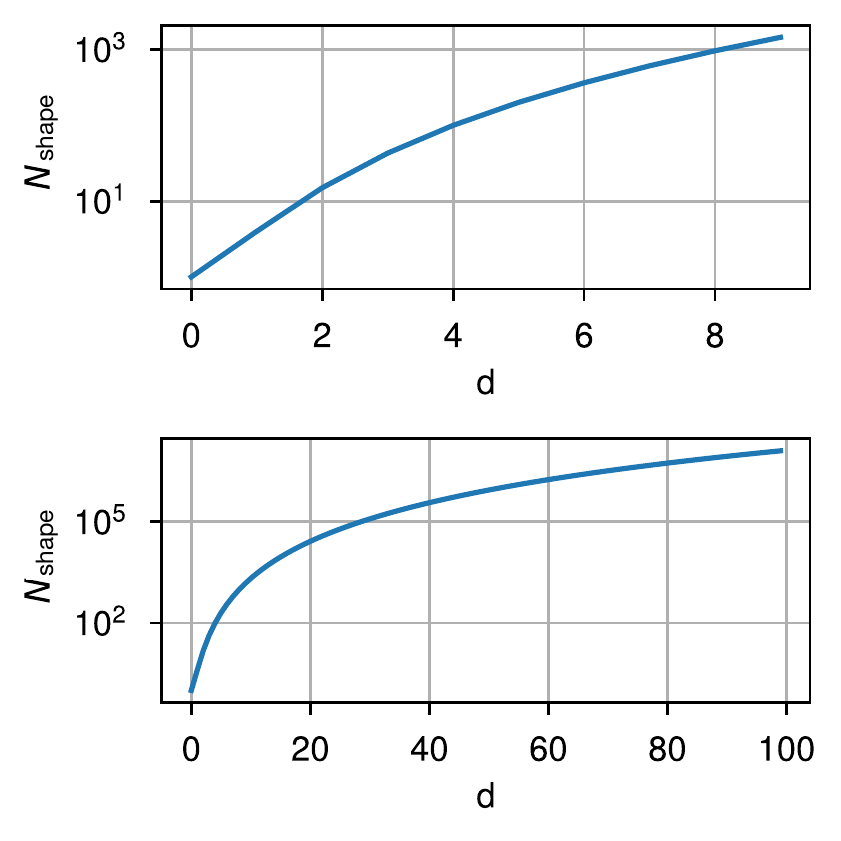}
\caption{Total number number of shape parameters 
$N_{\mathrm{shape}}$ as a function of the dimension $d$ over two different ranges.}
\end{figure}

The shape parameters entering the variational distribution in Eq.~(\ref{FinaLoss}) are 
\begin{align}
    &\mathrm{d.o.f.}\left[\hat{\bm{p}}\right] =  d,\label{dofLv}
\end{align}
\begin{align}
    &\mathrm{d.o.f.}\left[\mymat{L}^{\mathrm{v}}\right] =  \frac{d(d-1)}{2},\label{dofLv}\\
    &\mathrm{d.o.f.}\left[\bm{\Delta}^{\mathrm{v}}\right] = d,\\
    &\mathrm{d.o.f.}\left[\tilde{\mymat{Q}}^{\mathrm{v}}\right] =   \frac{d\left(2s - d - 1\right)}{2},\\
    &\mathrm{d.o.f.}\left[\mymat{L}^{\mathrm{m}}\right] =  \frac{s(s-1)}{2},\\
    &\mathrm{d.o.f.}\left[\bm{\Delta}^{\mathrm{m}}\right] = s.
\end{align}
Eqs.~(\ref{dofLv}) and (\ref{dofLv}) result from requiring the columns of $\mymat{L}^{\mathrm{v}}$ and $\mymat{L}^{\mathrm{m}}$ to have unitary norm.  Summing all up, the total number of shape parameters is 
\begin{equation}
    N_{\mathrm{shape}} = \frac{1}{8}\left(10d + 7d^2 + 6d^3 + d^4\right) . \label{totDOF}
\end{equation}
We show how Eq.~(\ref{totDOF}) scales with the dimension $d$ in Fig.~\ref{DimScaling}.

\section{QR-decomposition of rectangular matrices}
\label{App:qr}
We briefly describe the QR-decomposition for real-valued rectangular, rather than square matrices. 
The QR-decomposition factorizes an $m \times n$ matrix $\mymat{A}$ with $m > n$ into the product $\mymat{A} = \mymat{Q}\mymat{R}$, such that the matrix $\mymat{Q}$ is $m \times m$ and orthogonal, and the matrix $\mymat{R}$ is a rectangular $m\times n$ upper-triangular matrix. This implies the last $m-n$ rows of the matrix $\mymat{R}$ are zero. Square orthogonal matrices form a group, with the identity being $\mymat{I} = \mymat{Q}^T \mymat{Q}$.

The effect of the last rows in $\mymat{R}$ being zero is that many elements of $\mymat{Q}$ become irrelevant, as they will be multiplied by zero. This irrelevance can be denoted more explicitly by partitioning the matrices as
\begin{equation}
    \mymat{Q} = \begin{pmatrix}
    \mymat{Q}_1\ \mymat{Q}_2 \end{pmatrix}, \ \ \ \mymat{R} = \begin{pmatrix} \mymat{R_1}\\ \mymat{0}\end{pmatrix}.
\end{equation}
Here, $\mymat{R}_1$ is $n\times n$ upper triangular, and $\mymat{0}$ denotes the matrix of zero rows of $\mymat{R}$. The matrix $\mymat{Q}_2$ is $m\times (m-n)$ and denotes the irrelevant columns of $\mymat{Q}$ being multiplied by zero. $\mymat{Q}_1$ is a rectangular $m \times n$ matrix. 
We then have
\begin{equation}
    \mymat{A} = (\mymat{Q}_1\ \mymat{Q}_2) \begin{pmatrix} \mymat{R}_1 \\ \mymat{0}\end{pmatrix} = \mymat{Q}_1 \mymat{R}_1.
\end{equation}
For $m\times n$ orthogonal matrices such as $\mymat{Q}_1$, we have $\mymat{Q}_1^T\mymat{Q}_1 = \mymat{I}_{n\times n}$. In consequence, the matrix $\mymat{A}^T\mymat{A}$ will then have the decomposition $\mymat{A}^T\mymat{A} = \mymat{R}_1^T \mymat{Q}_1^T\mymat{Q}_1 \mymat{R}_1$, implying that $\mymat{R}_1$ is the Cholesky factor of $\mymat{A}^T\mymat{A}$ as the two orthogonal matrices yield the $n\times n$ identity matrix.

 % if your bibtex file is called example.bib

% Don't change these lines
\bsp	% typesetting comment
\label{lastpage}
\end{document}